\documentclass[aps,pre,twocolumn,floats,showpacs]{revtex4}

\usepackage{epsfig}
\usepackage{color}
\usepackage{bm}
\usepackage{latexsym}
\usepackage{times}

\begin{document}
\newcommand{\hide}[1]{}
\newcommand{\tbox}[1]{\mbox{\tiny #1}}
\newcommand{\half}{\mbox{\small $\frac{1}{2}$}}
\newcommand{\sinc}{\mbox{sinc}}
\newcommand{\const}{\mbox{const}}
\newcommand{\trc}{\mbox{trace}}
\newcommand{\intt}{\int\!\!\!\!\int }
\newcommand{\ointt}{\int\!\!\!\!\int\!\!\!\!\!\circ\ }
\newcommand{\eexp}{\mbox{e}^}
\newcommand{\bra}{\left\langle}
\newcommand{\ket}{\right\rangle}
\newcommand{\EPS} {\mbox{\LARGE $\epsilon$}}
\newcommand{\ar}{\mathsf r}
\newcommand{\im}{\mbox{Im}}
\newcommand{\re}{\mbox{Re}}
\newcommand{\bmsf}[1]{\bm{\mathsf{#1}}}
\newcommand{\mpg}[2][1.0\hsize]{\begin{minipage}[b]{#1}{#2}\end{minipage}}

\title{Scaling Properties of Multilayer Random Networks}

\author{J. A. M\'endez-Berm\'udez,$^1$ Guilherme Ferraz de Arruda,$^{2,3}$ Francisco A. Rodrigues,$^2$ and Yamir Moreno$^{3,4,5}$}

\affiliation{
$^1$Instituto de F\'{\i}sica, Benem\'erita Universidad Aut\'onoma de Puebla, 
Apartado Postal J-48, Puebla 72570, Mexico \\
$^2$Departamento de Matem\'{a}tica Aplicada e Estat\'{i}stica, Instituto de 
Ci\^{e}ncias Matem\'{a}ticas e de Computa\c{c}\~{a}o, Universidade de S\~{a}o 
Paulo - Campus de S\~{a}o Carlos, Caixa Postal 668, 13560-970 S\~{a}o Carlos, SP, Brazil \\
$^3$Institute for Biocomputation and Physics of Complex Systems (BIFI), University 
of Zaragoza, Zaragoza 50009, Spain \\
$^4$Department of Theoretical Physics, University of Zaragoza, Zaragoza 50009,
Spain \\
$^5$Complex Networks and Systems Lagrange Lab, Institute for Scientific 
Interchange, Turin, Italy
}

\date{\today}

\begin{abstract}
Multilayer networks are widespread in natural and manmade systems. Key properties of these networks are their spectral and eigenfunction characteristics, as they determine the critical properties of many dynamics occurring on top of them. In this paper, we numerically demonstrate that the 
normalized localization length $\beta$ of the eigenfunctions of multilayer random networks follows a simple scaling law given by $\beta=x^*/(1+x^*)$, with $x^*=\gamma(b_{\mbox{\tiny eff}}^2/L)^\delta$, $\gamma,\delta\sim 1$ and $b_{\mbox{\tiny eff}}$ being the effective 
bandwidth of the adjacency matrix of the network, whose size is $L=M\times N$. The reported scaling law for $\beta$ might help to better understand criticality in multilayer networks as well as used as to predict the eigenfunction localization properties of them. 
\end{abstract}


\maketitle

Real systems are naturally structured in levels or interconnected substructures, which in turn can 
be made of nodes organized in its own non-trivial manner~\cite{BoccalettiPR2014,Kivela2014}. 
For instance, individuals are connected according to friendship, working and family relations, defining 
different social circles, each of which can be thought of as a network. People and goods are transported through different mobility modes, such as airlines, roads and ships. All the previous systems made up what is nowadays referred to as multilayer networks ~\cite{Kivela2014}, which can also be found in biological and technological systems. Mathematically, the multilayer organization of real systems can be represented in different ways, being the matrix representation~\cite{Kivela2014,Cozzo2015} the most popular. 

On the other hand, it has also been shown that many critical properties of several phenomena that take place on top of complex networks are determined by the topology of them, and specifically by the spectral and eigenfunction properties of the adjacency and the Laplacian matrices of the networks. One particularly suitable approach to address the relation between the structure and the dynamics of a networked system is given by Random Matrix Theory 
(RMT), see for instance~\cite{Bandyopadhyay,MAM15}. Random Matrix Theory has numerous 
applications in many different fields, from condensed matter physics to financial markets~\cite{RMT}. 
In the case of complex networks, the use of RMT techniques might reveal universal properties, i.e., the 
nearest neighbor spacing distribution of the eigenvalues of the adjacency matrices of various model 
networks follow Gaussian Orthogonal Ensemble (GOE) statistics~\cite{Bandyopadhyay}. For instance, the 
analysis of Erd\"os-R\'enyi networks shows that the level spacing distribution and the entropic 
eigenfunction localization length of the adjacency matrices are universal for fixed average 
degrees~\cite{MAM15}.

Universal properties are always of interest, as they allow to reduce the set of parameters describing the system and provide relations that allow to deduce its behavior from those few global parameters, without the need of having precise details of the system. In this paper, we study whether there are universal scaling properties in multilayer systems. To this end, we perform a scaling analysis of the eigenfunction localization properties of multilayer networks using RMT models and techniques. We explore multilayer networks whose networks of layers are of two types: (i) a line and (ii) a complete graph (node-aligned multiplex networks). In the first case, we study weighted layers coupled by 
weighted matrices, whereas in the latter case, weighted and binary layers coupled by identity matrices are considered. We demonstrate that the normalized localization length of the eigenfunctions of multilayer random networks exhibits a well defined scaling function, being the scaling law robust for all the aforementioned networks. Our results can be used to predict or design the localization features of the eigenfunctions of multilayer random networks and to better understand critical properties that depend on eigenfunction properties.

\begin{figure}[t]
\centerline{\includegraphics[width=\columnwidth]{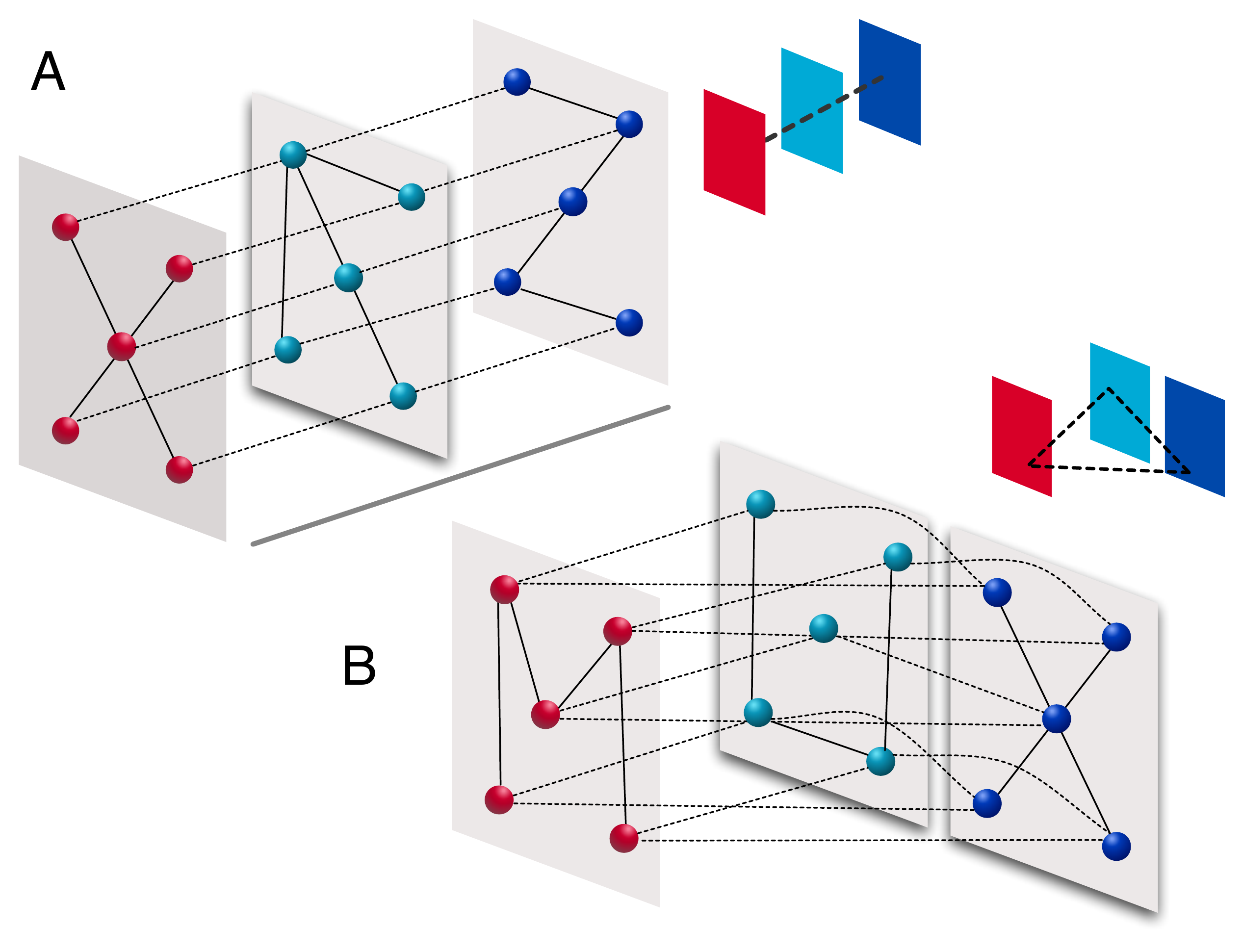}}
\caption{(color online) Illustration of the two types of multilayer networks studied here. The network of layers are 
(a) a line and (b) a complete network. Here, each network is composed by $M=3$
layers having $N=5$ nodes.}
\label{fig:schematic}
\end{figure}

A multilayer network is formed by $M$ undirected random layers with
corresponding adjacency matrices $A^{(m)}$ having $N_m$ nodes each. The respective adjacency 
matrix of the whole network is expressed by 
${\bf A} = \bigoplus_{m=1}^M A^{(m)} + {\it p} \bf{C}$, where $\bigoplus$ represents the Kronecker
product, $p$ is a parameter that defines the strength of the inter-layer edges and $\bf{C}$  is the 
interlayer coupling matrix, whose elements represent the relations between nodes in different layers. 
Observe that the coupling matrix has implicitly the information of a network of layers. On such network, 
nodes represent layers of the multilayer network, i.e., there is an edge if at least one node is connected 
on both layers. Moreover, the network of layers can be extracted in different manners, as can be seen 
in~\cite{Garcia2014}. Examples of multilayers are shown in Fig.~\ref{fig:schematic}. 
Observe that the spectra of the adjacency matrix  {\bf A} is a function of the parameter $p$. As a 
consequence, eigenvalue crossings, structural transitions~\cite{Radicchi2013}, near 
crossings~\cite{Arruda2015} or localization problems~\cite{Goltsev2012, Arruda2015} are inherent to 
the network spectra, depending on $p$ for multilayer networks. Regarding dynamical processes, such 
parameter plays a fundamental role. For instance on diffusion processes it can drive the network to what 
is called super-diffusion~\cite{Gomez2013}, which means that the time scale of the multiplex is faster 
than the time scale observed when the layers are separated. Another important example is the case of contagion dynamics, for which $p$ can be interpreted as the ratio of intra and inter-layer 
spreading rates, giving raise to the existence of both localized and delocalized 
states~\cite{Arruda2015}. Here we restrict ourselves to $p=1$: for $p \ll 1$ the layers can be considered as uncoupled, while for $p \gg 1$ the topology of the network of layers dominates the spectral properties~\cite{Garcia2014,Cozzo2016}. In this 
way, $p = 1$ represents a suitable intermediary case (multilayer phase). 

We define two ensembles of multilayer random networks as adjacency matrices. As the first model 
we consider a network of layers on a line, see Fig.~\ref{fig:schematic}(a), whose adjacency matrix 
${\bf A}$ has the form 
\begin{equation}
{\bf A} =
\left(
\begin{array}{ccccc}
A^{(1)} & C^{(1,2)} & 0 & \cdots & 0 \\
C^{(2,1)} & A^{(2)} & C^{(2,3)} & \ & 0 \\
0 & C^{(3,2)} & A^{(3)} & \ & 0 \\
\vdots & \ & \ & \ddots & C^{(M-1,M)} \\
0 & 0 & 0 & C^{(M,M-1)} & A^{(M)}
\end{array}
\right) \ ,
\label{eq:A_line}
\end{equation}
where $\left(C^{(m,m')}\right)_{i,j}=\left(C^{(m,m')}\right)^{\mbox{\tiny T}}_{j,i}$ are real rectangular 
matrices of size $N_m \times N_{m'}$ and 0 represents null matrices. Furthermore, we consider a 
special class of matrices $A^{(m)}$ and $C^{(m,m')}$ which are characterized by the sparsities 
$\alpha_A$ and $\alpha_C$, respectively. In other words, since with a probability $\alpha_*$ their 
elements can be removed, these matrices represent Erd\"os-R\'enyi--type random networks. Notice 
that when the $N_m$ are all the same $N_m=\mbox{constant}\equiv N$, which is the case we 
explore  here. Also, the adjacency matrix ${\bf A}$ has the structure of a block-banded matrix of 
size $L=M\times N$. In addition, we consider this model as a model of weighted networks; i.e., the non-vanishing elements ${\bf A}_{i,j}$ are independent Gaussian variables with 
zero mean and variance $1+\delta_{i,j}$. 
We justify the addition of self-loops and random weights to edges by recognizing
that in real-world networks the nodes and the interactions between them are in general
non-equivalent. Moreover, with this prescription we retrieve well known random 
matrices \citep{Metha} in the appropriate limits: According to this definition a 
diagonal random matrix is obtained for $\alpha_A=\alpha_C=0$ (Poisson case), whereas 
the GOE is recovered when $\alpha_A=\alpha_C=1$ and 
$M=2$. For simplicity, and without loss 
of generality, in this work we consider the case where $\alpha\equiv\alpha_A=\alpha_C$.
As an example, this network model can be applied to transportation networks, where the
inter-layer edges represent connections between two different means of transport.  An obvious constraint is that no layer can be connected to more than two layers. In addition to the above configuration, we are also interested on the node-aligned multiplex
case, whose network of layers is a complete graph~\cite{SM}, see Fig.~\ref{fig:schematic}(b).

There is a well known RMT model known as the banded random matrix (BRM) model which
was originally introduced to emulate quasi-one-dimensional 
disordered wires of length $L$ and width $b$ (with $b\ll L$). The BRM ensemble is defined 
as the set of $L\times L$ real symmetric matrices whose entries are independent Gaussian 
random variables with zero mean and variance $1+\delta_{i,j}$ if $|i-j|<b$ and zero
otherwise. Therefore $b$ is the number of nonzero elements in the first matrix row which
equals 1 for diagonal, 2 for tridiagonal, and $L$ for matrices of the GOE. There are several 
numerical and theoretical studies available for this model, see for example 
Refs.~\cite{CMI90,EE90,FM91,CIM91,FM92,MF93,FM93,FM94,I95,MF96,CGM97,S97,KPI98,KIP99,W02}. 
In particular, outstandingly, it has been found
\cite{CMI90,EE90,FM91,FM92,MF93,FM93,FM94,I95} 
that the eigenfunction properties of the BRM model, characterized by the {\it scaled localization 
length} $\beta$ (to be defined below), are {\it universal} for the fixed ratio $x = b^2/L$.
More specifically, it was numerically and theoretically shown that the scaling function
\begin{equation}
\beta = \gamma x/(1+\gamma x) \ ,
\label{betascaling0}
\end{equation}
with $\gamma\sim 1$, holds for the eigenfunctions of the BRM model \footnote{It is relevant 
to mention that the scaling (\ref{betascaling0}) was also shown to
be valid for the kicked-rotator model \cite{CGIS90} (a {\it quantum-chaotic} 
system characterized by a random-like banded Hamiltonian matrix), the
one-dimensional Anderson model, and the Lloyd model \cite{CGIFM92}.}.

Admittedly, the ensemble of adjacency matrices of the multilayer network with
layers on a line, see Eq.~(\ref{eq:A_line}), can be considered as a {\it non-homogeneous 
diluted version} of the BRM model. Therefore, motivated by the similarity between these 
two complex matrix models, we propose the study of eigenfunction properties of the 
adjacency matrices of multilayer and multiplex random networks as a function of the parameter
\begin{equation}
x = b_{\mbox{\tiny eff}}^2/L \ ,
\label{x}
\end{equation}
where $b_{\mbox{\tiny eff}}\equiv b_{\mbox{\tiny eff}}(N,\alpha)$ is the adjacency matrix 
effective bandwidth and $L=M\times N$. 

A commonly accepted tool to characterize quantitatively the complexity of the 
eigenfunctions of random matrices (and of Hamiltonians corresponding to disordered 
and quantized chaotic systems) is the information or Shannon entropy $S$. This measure 
provides the number of principal components of an eigenfunction in a given basis. 
In fact, $S$ has been already used to characterize the eigenfunctions of the adjacency 
matrices of random network models; see some examples in 
Refs.~\cite{ZYYL08,GT06,JSVL10,PS12,MRP14,MAM15}.
The Shannon entropy for the eigenfunction $\Psi^l$ is given as
$S = -\sum_{n=1}^L (\Psi^l_n)^2 \ln (\Psi^l_n)^2$ and allows computing the scaled localization length as
\cite{I90}
\begin{equation}
\label{beta}
\beta = \exp\left( \bra S \ket - S_{\tbox{GOE}} \right) \ ,
\end{equation}
where $S_{\tbox{GOE}}\approx\ln(L/2.07)$, which is used as a reference, 
is the entropy of a random eigenfunction 
with Gaussian distributed amplitudes (i.e.,~an eigenfunction of the GOE). 
With this definition for $\beta$ and in the case of the multilayer network with
layers on a line,
when $\alpha=0$ (i.e.,~when all vertices in the network are isolated), since the 
eigenfunctions of the adjacency matrices of Eq.~(\ref{eq:A_line}) have only one 
non-vanishing component with magnitude equal to one, $\bra S \ket=0$ and 
$\beta\approx 2.07/L$. On the other hand, when all nodes in this multilayer network are 
fully connected we recover the GOE and $\bra S \ket=S_{\tbox{GOE}}$. Thus, the {\it fully chaotic} eigenfunctions extend over the $L$ available vertices in the network
and $\beta=1$. Therefore, $\beta$ can take values in the range $(0,1]$.

Here, as well as in BRM model studies, we look for the scaling properties of the 
eigenfunctions of our multilayer random network models through $\beta$.
Below we use exact numerical diagonalization to obtain eigenfunctions 
$\Psi^l$ ($l=1\ldots L$) of the adjacency matrices of large ensembles of multilayer 
random networks characterized by $M$, $N$, and $\alpha$. 
We perform the average $\bra S \ket$ taking half of the eigenfunctions, around the 
band center, of each adjacency matrix.

\begin{figure}[t]
\centering
\centerline{\includegraphics[width=\columnwidth]{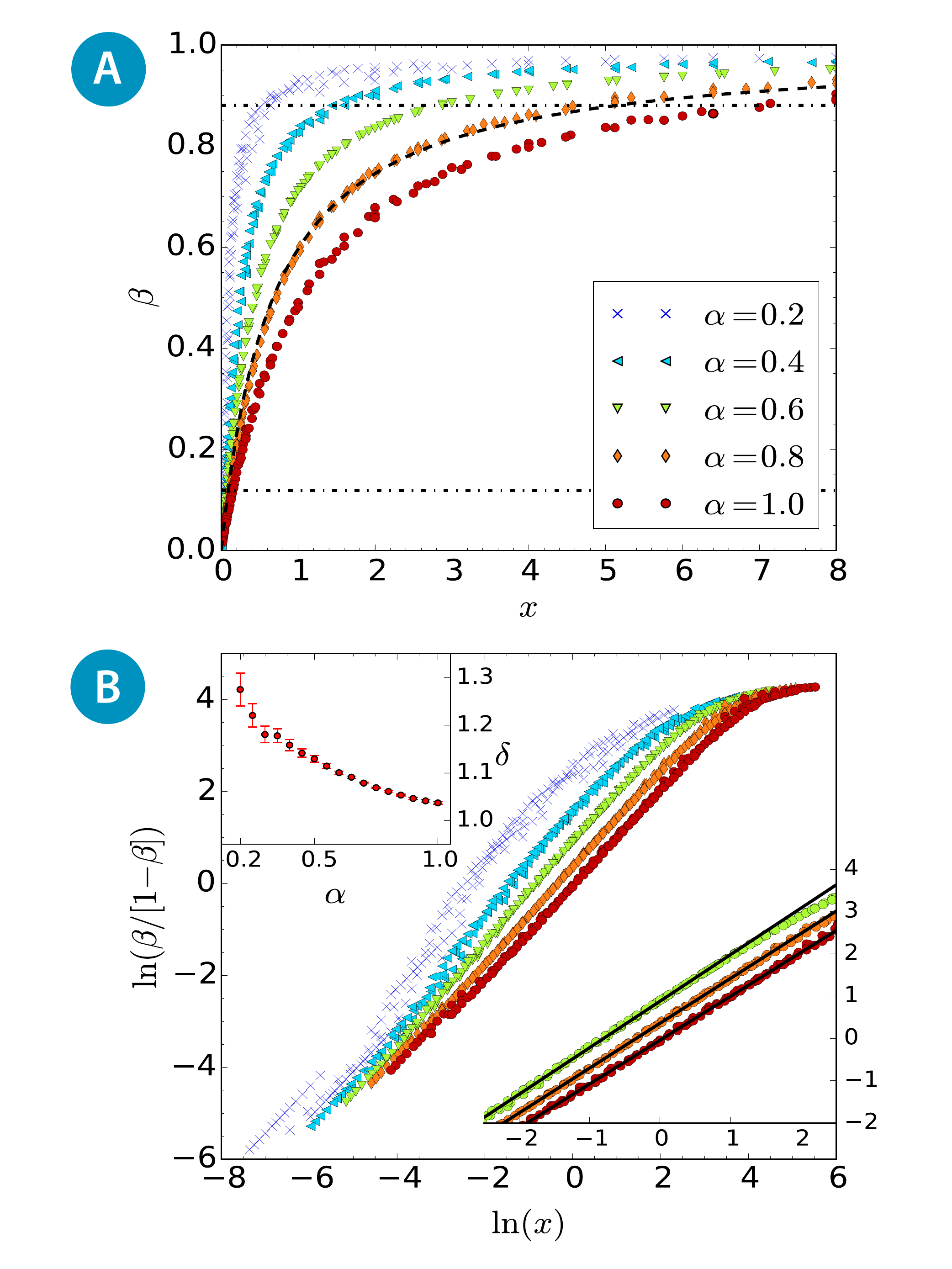}}
\caption{(Color online) 
(a) Scaled localization length $\beta$ as a function of $x=b_{\mbox{\tiny eff}}^2/L$ 
for ensembles of multilayer networks characterized by the sparsity $\alpha$. 
The black dashed line close to the data for $\alpha=0.8$ corresponds to 
Eq.~(\ref{betascaling0}) with $\gamma=1.4$. Horizontal black dot-dashed lines at 
$\beta\approx 0.12$ and 0.88 are shown as a reference, see the text.
(b) Logarithm of $\beta/(1-\beta)$ as a function of $\ln(x)$. 
Upper inset: Power $\delta$, from the fittings of the data with Eq.~(\ref{betascaling2}), 
as a function of $\alpha$. 
Lower inset: Enlargement in the range $\ln[\beta/(1-\beta)]=[-2,4]$ including data for 
$\alpha=0.6$, 0.8, and 1. Lines are fittings of the data using Eq.~(\ref{betascaling2}).}
\label{fig:BBRM}
\end{figure}
\begin{figure}
\centerline{\includegraphics[width=\columnwidth]{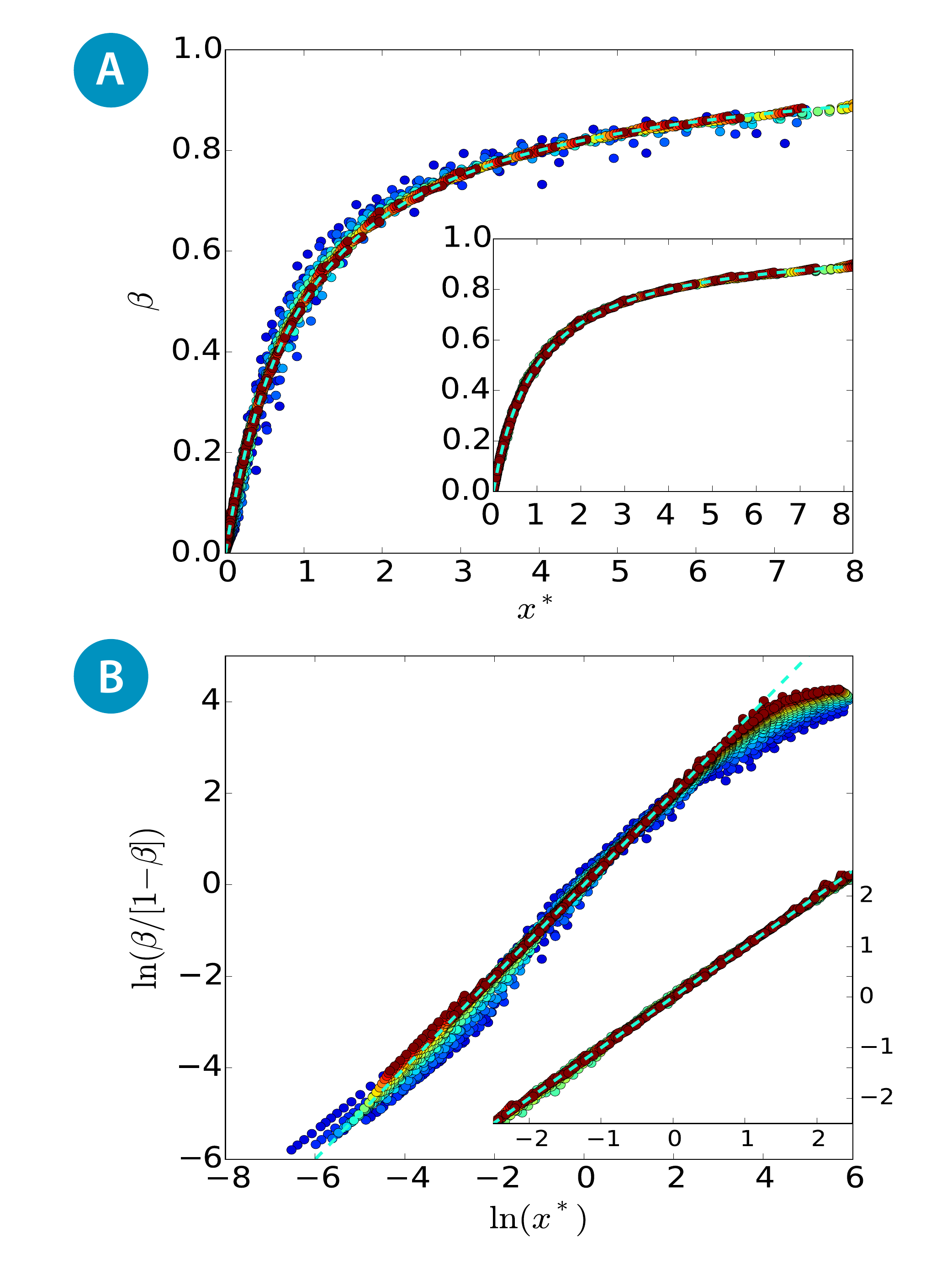}}
\caption{(Color online) 
(a) $\beta$ as a function of $x^*$ [as defined in Eq.~(\ref{x*})] for ensembles of multilayer 
networks with $\alpha\in [0.2,1]$ in steps of $0.05$.
Inset: Data for $\alpha\in [0.5,1]$ in steps of $0.05$. Orange dashed lines in main 
panel and inset are Eq.~(\ref{betax*}).
(a) Logarithm of $\beta/(1-\beta)$ as a function of $\ln(x^*)$ for 
$\alpha\in [0.2,1]$ in steps of $0.05$. 
Inset: Enlargement in the range $\ln[\beta/(1-\beta)]=[-2,2]$ including curves for 
$\alpha\in [0.5,1]$ in steps of $0.05$. Orange dashed lines in main panel and inset 
are Eq.~(\ref{betascaling3}).}
\label{fig:BBRM_collapsed}
\end{figure}

We now analyze in detail the multilayer network model with adjacency matrix given by 
Eq.~(\ref{eq:A_line}). In Fig.~\ref{fig:BBRM}(a), we present $\beta$ as a function of $x$, see Eq.~(\ref{x}),
for ensembles of networks characterized by the sparsity $\alpha$. We have 
defined $b_{\mbox{\tiny eff}}$ as the average number of non-vanishing elements per 
adjacency-matrix row, i.e.,
\begin{equation}
b_{\mbox{\tiny eff}} = 2N\alpha \ .
\label{beff}
\end{equation}

We observe that the curves of $\beta$ vs.~$x$ in Fig.~\ref{fig:BBRM}(a) have a functional 
form similar to that for the BRM model. To show this we are including 
Eq.~(\ref{betascaling0}) 
(black dashed line) with $\gamma=1.4$ (the value of $\gamma$ reported in 
Ref.~\cite{CMI90} for the BRM model) which is very close to our data for $\alpha=0.8$. 
In addition, in Fig.~\ref{fig:BBRM}(b) the logarithm of 
$\beta/(1-\beta)$ as a function of $\ln(x)$ is presented. 
The quantity $\beta/(1-\beta)$ was useful in the study of the scaling properties of the BRM 
model \cite{CMI90,FM92} because $\beta/(1-\beta) = \gamma x$, which is equivalent to scaling
(\ref{betascaling0}), implies that a plot of $\ln[\beta/(1-\beta)]$  vs.~$\ln(x)$ is a straight line 
with unit slope. Even though, this statement is valid for the BRM model in a wide range of 
parameters (i.e.,~for $\ln[\beta/(1-\beta)]<2$) it does not apply to our multilayer random 
network model; see Fig.~\ref{fig:BBRM}(b). In fact, from this figure we observe that plots of 
$\ln[\beta/(1-\beta)]$ vs.~$\ln(x)$ are straight lines (in a wide range of $x$) with 
a slope that depends on the sparsity $\alpha$. Therefore, we propose the scaling law 
\begin{equation}
\beta/(1-\beta) = \gamma x^\delta \ ,
\label{betascaling2}
\end{equation}
where both $\gamma$ and $\delta$ depend on $\alpha$. Indeed, Eq.~(\ref{betascaling2})
describes well our data, mainly in the range $\ln[\beta/(1-\beta)]=[-2,2]$, as can be 
seen in the inset of Fig.~\ref{fig:BBRM}(b) where we show the numerical data for 
$\alpha=0.6$, 0.8 and 1 and include fittings through Eq.~(\ref{betascaling2}).
We stress that the range $\ln[\beta/(1-\beta)]=[-2,2]$ corresponds to a
reasonable large range of $\beta$ values, $\beta\approx[0.12,0.88]$, whose bounds
are indicated with horizontal dot-dashed lines in Fig.~\ref{fig:BBRM}(a).
Finally, we notice that the power $\delta$, obtained from the fittings of the data 
using Eq.~(\ref{betascaling2}), is very close to unity for all the sparsity
values we consider here (see the upper inset of Fig.~\ref{fig:BBRM}(b)).

Therefore, from the analysis of the data in Fig.~\ref{fig:BBRM}, we are able to write down 
a {\it universal scaling function} for the scaled localization length $\beta$ of the
eigenfunctions of our multilayer random network model as
\begin{equation}
\beta/(1-\beta) = x^* \ ,
\label{betascaling3}
\end{equation}
where the scaling parameter $x^*=\gamma x^\delta$, as a function of the multilayer network 
parameters, is given by
\begin{equation}
x^* \equiv \gamma \left( 4N\alpha^2/M \right)^\delta \ .
\label{x*}
\end{equation}
To validate Eq.~(\ref{betascaling3}) in Fig.~\ref{fig:BBRM_collapsed}(b) we present 
again the
data for $\ln[\beta/(1-\beta)]$ shown in Fig.~\ref{fig:BBRM}(b) but now as a function 
of $\ln(x^*)$. We do observe that curves for different values of $\alpha$ fall on 
top of Eq.~(\ref{betascaling3}) for a wide range of the variable $x^*$.
Moreover, the collapse of the numerical data is excellent in the range $\ln[\beta/(1-\beta)]=[-2,2]$ for $\alpha\ge 0.5$, as
shown in the inset of Fig.~\ref{fig:BBRM_collapsed}(b).

Finally, we rewrite Eq.~(\ref{betascaling3}) into the equivalent, but explicit, 
scaling function for $\beta$:
\begin{equation}
\beta = x^*/(1+x^*) \ .
\label{betax*}
\end{equation}
In Fig.~\ref{fig:BBRM_collapsed}(a) we confirm the validity of Eq.~(\ref{betax*}).
We emphasize that the universal scaling given in Eq.~(\ref{betax*})
extends outside the range $\beta\approx[0.12,0.88]$, for which Eq.~(\ref{betascaling2})
was shown to be valid, see the main panel of Fig.~\ref{fig:BBRM_collapsed}(a). Furthermore, 
the collapse of the numerical data following Eq.~(\ref{betax*}) is remarkably good for 
$\alpha\ge 0.5$, as shown in the inset of Fig.~\ref{fig:BBRM_collapsed}(a). Additionally, we have verified~\cite{SM} that the scaling (\ref{betax*}) is also applicable to node-aligned multiplex networks $-$which are relevant for certain applications$-$, once the effective bandwidth $b_{\mbox{\tiny eff}}$ is properly defined. 

Summarizing, in this study we have demonstrated that the normalized localization length $\beta$ of the
eigenfunctions of multilayer random networks scales as $x^*/(1+x^*)$. Here, $x^*=\gamma(b_{\mbox{\tiny eff}}^2/L)^\delta$; where $b_{\mbox{\tiny eff}}$ is the effective bandwidth of the network's adjacency matrix, $L$ is the adjacency matrix size,
and $\gamma,\delta\sim 1$. We showed that such scaling law is robust covering weighted multilayer
and both weighted and unweighted node-aligned multiplex networks~\cite{SM}. Our results might shed additional light on the critical properties and structural organization of multilayer systems. Interestingly enough, our findings might be used to either predict or design (e.g, tune), by means of Eq.~(\ref{betax*}), the localization properties of the eigenfunctions of multilayer random 
networks. For instance, we anticipate the following cases: (i) Due to the banded nature of the adjacency matrices of the network models considered here, $b_{\mbox{\tiny eff}}<L$, it is unlikely to observe fully delocalized eigenfuctions unless the value 
of $x^*$ is driven to large values, for example, by increasing the size of the subnetworks $N$ and/or their sparsity $\alpha$ for a fixed value of $M$; (ii) For a fixed subnetwork size $N$ and sparsity $\alpha$, the eigenfunctions of the multilayer network become more localized when increasing the number of subnetworks $M$; and (iii) the procedure of adding/removing subnetworks in our network models may be used to tune their conduction properties~\footnote{In order to study transport properties one may consider to open the 
multilayer random network model by attaching conducting leads to chosen vertices, see for example \cite{MAM13,M16}.}, 
since $M$ could drive the network from a regime of delocalized eigenfunctions (metallic regime), $x^*\gg 1$, to a regime of localized eigenfunctions (insulating regime),  $x^*\ll 1$. We hope our results motivate further numerical and theoretical studies. \\

{\it Acknowledgments}.--
This work was partially supported by VIEP-BUAP (Grant No.~MEBJ-EXC16-I), 
Fondo Institucional PIFCA (Grant No.~BUAP-CA-169), and 
CONACyT (Grant Nos.~I0010-2014/246246 and CB-2013/220624).
FAR acknowledges CNPq (Grant No.~305940/2010-4),
FAPESP (Grant No.~2011/50761-2 and 2013/26416-9), and 
NAP eScience - PRP - USP for financial support. 
GFA would like to acknowledge FAPESP (grants 2012/25219-2 and 2015/07463-1) 
for the scholarship provided. Y. M. acknowledges support from the Government of Arag\'on, Spain through a grant to the group FENOL, by MINECO and FEDER funds (grant FIS2014-55867-P) and by the European Commission FET-Proactive Project Multiplex (grant 317532).

\appendix

\section{Scaling analysis of Multiplex Networks}

In the node-aligned multiplex case, whose network of layers is a complete graph,
the coupling matrices are 
restricted to identity matrices and all layers have the same number of nodes, see 
Fig.~\ref{fig:schematic}(b). The adjacency matrix of a node-aligned multiplex is given as
\begin{equation}
{\bf A} =
\left(
\begin{array}{ccccc}
A^{(1)} & I & I & \cdots & I \\
I & A^{(2)} & I & \ & I \\
I & I & A^{(3)} & \ & I \\
\vdots & \ & \ & \ddots & I \\
I & I & I & I & A^{(M)}
\end{array}
\right) \ .
\label{eq:A_mux}
\end{equation}
Similarly to the multilayer model of Eq.~(\ref{eq:A_line}), this configuration is characterized by 
the sparsity $\alpha$
which we choose to be constant for all the $M$ matrices $A^{(m)}$ of size $N\times N$ 
composing the adjacency matrix ${\bf A}$ of size $L=M\times N$. 
Additionally, the configuration (\ref{eq:A_mux}) is considered in two different setups: 
weighted and unweighted multiplex without self-loops. In the weighted case, the non-vanishing elements of the matrices $A^{(m)}$ are chosen as independent Gaussian variables with zero mean and variance $1+\delta_{i,j}$. A realistic example of this 
configuration are online social systems, where each layer represents a different online network (e.g., Facebook, Twitter and Google+, etc). In the unweighted case, the non-vanishing elements of $A^{(m)}$ are equal to unity. In (\ref{eq:A_mux}), $I$ are identity matrices of size $N\times N$.

\begin{figure}[t]
\centering
\centerline{\includegraphics[width=\columnwidth]{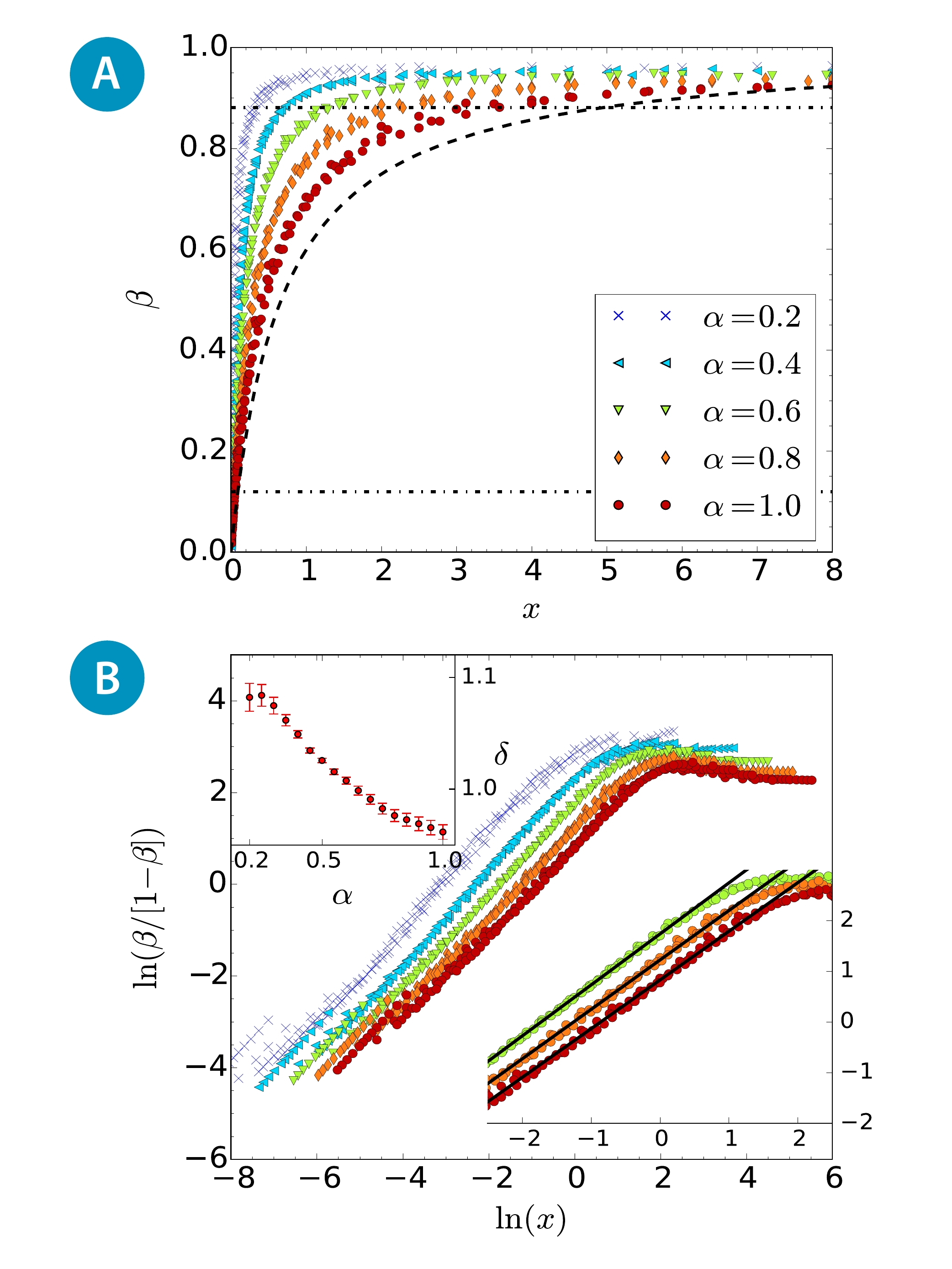}}
\caption{(Color online) 
(a) Scaled localization length $\beta$ as a function of $x=b_{\mbox{\tiny eff}}^2/L$ 
for ensembles of weighted multiplex networks characterized by the sparsity $\alpha$. 
The black dashed line close to the data for $\alpha=0.6$ corresponds to 
Eq.~(\ref{betascaling0}) with $\gamma=1.4$. Horizontal black dot-dashed lines at 
$\beta\approx 0.12$ and 0.88 are shown as a reference, see the text.
(b) Logarithm of $\beta/(1-\beta)$ as a function of $\ln(x)$. 
Upper inset: Power $\delta$, from the fittings of the data with Eq.~(\ref{betascaling2}), 
as a function of $\alpha$. 
Lower inset: Enlargement in the range $\ln[\beta/(1-\beta)]=[-2,2]$ including data for 
$\alpha=0.6$, 0.8, and 1. Lines are fittings of the data with Eq.~(\ref{betascaling2}).}
\label{fig:W_Mux}
\end{figure}
\begin{figure}[t]
\centerline{\includegraphics[width=\columnwidth]{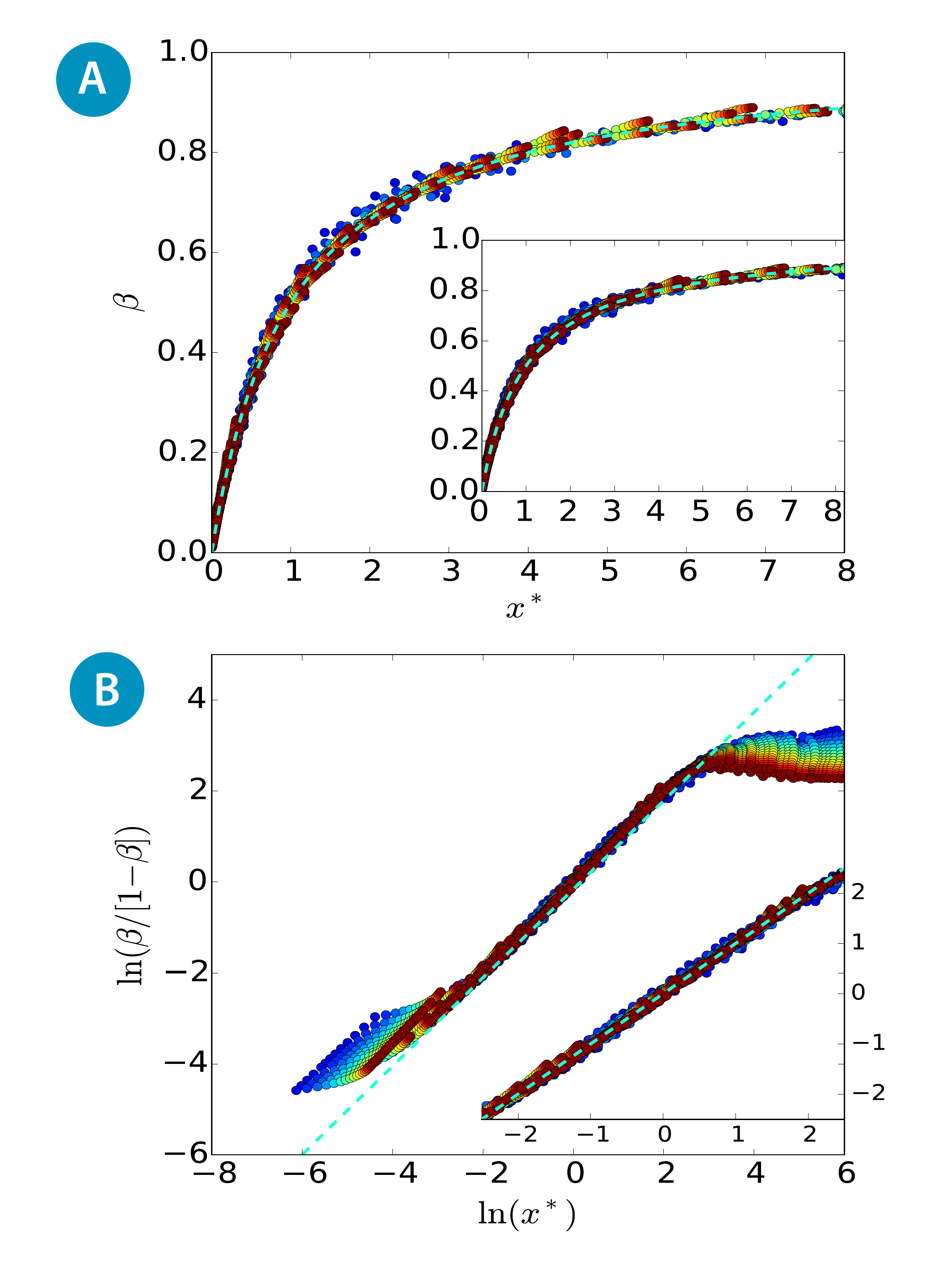}}
\caption{(Color online) 
(a) $\beta$ as a function of $x^*$ [as defined in Eq.~(\ref{x*})] for ensembles of weighted 
multiplex networks with $\alpha\in [0.2,1]$ in steps of $0.05$.
Inset: Data for $\alpha\in [0.5,1]$ in steps of $0.05$. Orange dashed lines in main 
panel and inset are Eq.~(\ref{betax*}).
(a) Logarithm of $\beta/(1-\beta)$ as a function of $\ln(x^*)$ for 
$\alpha\in [0.2,1]$ in steps of $0.05$. 
Inset: Enlargement in the range $\ln[\beta/(1-\beta)]=[-2,2]$ including curves for 
$\alpha\in [0.5,1]$ in steps of $0.05$. Orange dashed lines in main panel and inset 
are Eq.~(\ref{betascaling3}).}
\label{fig:W_Mux_collapsed}
\end{figure}

\subsection{Weigthed Multiplex}

\begin{figure}[t]
\centering
\centerline{\includegraphics[width=\columnwidth]{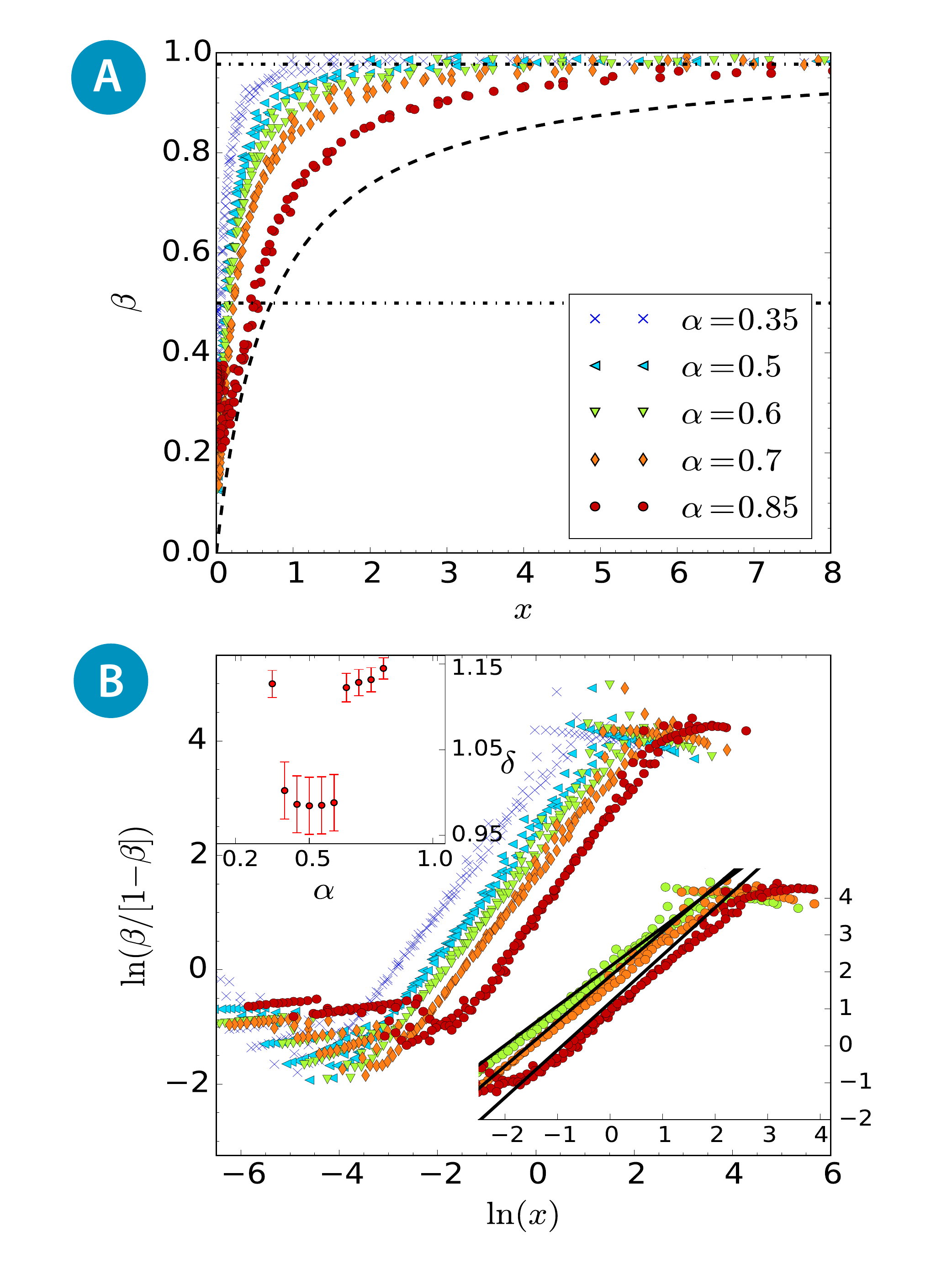}}
\caption{(Color online) 
(a) Scaled localization length $\beta$ as a function of $x=b_{\mbox{\tiny eff}}^2/L$ 
for ensembles of unweighted multiplex networks characterized by the sparsity $\alpha$. 
The black dashed line corresponds to 
Eq.~(\ref{betascaling0}) with $\gamma=1.4$. Horizontal black dot-dashed lines at 
$\beta\approx 0.5$ and 0.98 are shown as a reference, see the text.
(b) Logarithm of $\beta/(1-\beta)$ as a function of $\ln(x)$. 
Upper inset: Power $\delta$, from the fittings of the data with Eq.~(\ref{betascaling2}), 
as a function of $\alpha$. 
Lower inset: Enlargement in the range $\ln[\beta/(1-\beta)]=[-2,4]$ including data for 
$\alpha=0.6$, 0.7, and 0.85. Lines are fittings of the data with Eq.~(\ref{betascaling2}).}
\label{fig:A_Mux}
\end{figure}
\begin{figure}[t]
\centerline{\includegraphics[width=\columnwidth]{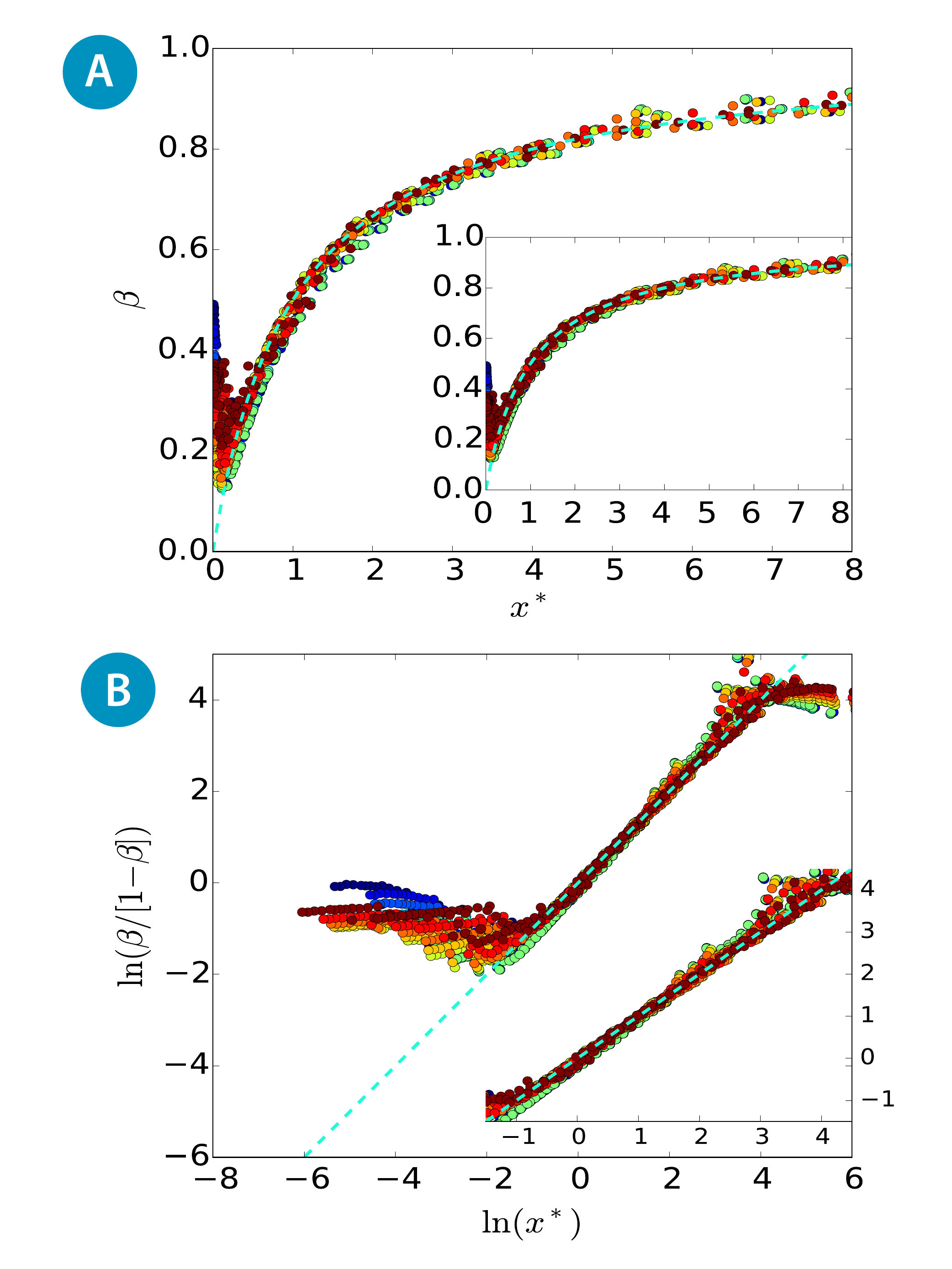}}
\caption{(Color online) 
(a) $\beta$ as a function of $x^*$ [as defined in Eq.~(\ref{x*})] for ensembles of unweighted 
multiplex networks with $\alpha\in [0.2,1]$ in steps of $0.05$.
Inset: Data for $\alpha\in [0.5,1]$ in steps of $0.05$. Orange dashed lines in main 
panel and inset are Eq.~(\ref{betax*}).
(a) Logarithm of $\beta/(1-\beta)$ as a function of $\ln(x^*)$ for 
$\alpha\in [0.2,1]$ in steps of $0.05$. 
Inset: Enlargement in the range $\ln[\beta/(1-\beta)]=[-1,4]$ including curves for 
$\alpha\in [0.5,1]$ in steps of $0.05$. Orange dashed lines in main panel and inset 
are Eq.~(\ref{betascaling3}).}
\label{fig:A_Mux_collapsed}
\end{figure}

Now we consider weighted multiplex networks (i.e., where the non-vanishing elements of the 
adjacency matrices $A^{(m)}$ in (\ref{eq:A_mux}) are chosen as independent Gaussian variables 
with zero  mean and variance $1+\delta_{i,j}$). We follow the same methodology as in the multilayer 
case. Thus, in Fig.~\ref{fig:W_Mux}(a) we first present curves of $\beta$ vs.~$x$ as given in 
Eqs.~(\ref{beta}) and (\ref{x}), respectively; however, we redefine
$b_{\mbox{\tiny eff}}$ as 
\begin{equation}
b_{\mbox{\tiny eff}} = N\alpha \ ,
\label{beff2}
\end{equation}
which is the average number of non-vanishing elements per row inside the 
adjacency-matrix band in the multiplex setup. 
From Fig.~\ref{fig:W_Mux}(a) we observe that the curves of $\beta$ vs.~$x$ have
functional forms similar to those for the multilayer model (compare with Fig.~\ref{fig:BBRM}(a));
however, with
larger values of $\beta$ for given values of $x$. As a reference we also include 
Eq.~(\ref{betascaling0}) (black-dashed line) with $\gamma=1.4$, corresponding to the BRM 
model, which is even below  
the data for $\alpha=1$. Moreover, in Fig.~\ref{fig:W_Mux}(b) we show the logarithm of $\beta/(1-\beta)$ as a function of $\ln(x)$. As in the multilayer case, here we observe that 
plots of $\ln[\beta/(1-\beta)]$ vs.~$\ln(x)$ are straight lines mainly in the range
$\ln[\beta/(1-\beta)]=[-2,2]$ with a slope that depends on the sparsity $\alpha$.
We indicate the bounds of this range with horizontal dot-dashed lines in Fig.~\ref{fig:W_Mux}(a).
Therefore, the scaling law of Eq.~(\ref{betascaling2}) is also valid here. Indeed, in the upper inset of 
Fig.~\ref{fig:W_Mux}(b) we report the power $\delta$ obtained from fittings of the data with 
Eq.~(\ref{betascaling2}).

In order to validate the scaling hypothesis of Eq.~(\ref{betascaling2}) for the node-aligned multiplex setup, in 
Fig.~\ref{fig:W_Mux_collapsed}(b) we present the data for $\ln[\beta/(1-\beta)]$ shown in 
Fig.~\ref{fig:W_Mux}(b), but now as a function of $\ln(x^*)$. We observe that curves for different 
values of  $\alpha$ fall on top of Eq.~(\ref{betascaling3}) for a wide range of the variable $x^*$. 
Moreover, the collapse of the numerical data on top of 
Eq.~(\ref{betascaling3}) is excellent in the range $\ln[\beta/(1-\beta)]=[-2,2]$ for $\alpha\ge 0.5$, as 
shown in the inset of Fig.~\ref{fig:W_Mux_collapsed}(b).
Finally, in Fig.~\ref{fig:W_Mux_collapsed}(a) we confirm the validity of Eq.~(\ref{betax*}) which is 
as good here as for the multilayer case. We emphasize that 
the collapse of the numerical data on top of Eq.~(\ref{betax*}) is remarkably good for 
$\alpha\ge 0.5$, as shown in the inset of Fig.~\ref{fig:W_Mux_collapsed}(a).

\subsection{Unweighted Multiplex}

The last analyzed scenario is the binary multiplex case. We recall that, in contrast to the two 
previous random network models, this model does not include weighted self-loops. Therefore the Poisson limit is not recovered when $\alpha\to 0$ and $\beta$ is not well defined there. Thus, we 
will compute $\beta$ for values of $x$ as smaller as the adjacency-matrix diagonalization produces 
meaningful results. Also, as for the weighted multiplex, we use here the effective bandwidth given in
Eq.~(\ref{beff2}).

The conducted experiments are similar to the previous ones. Then, in Figs.~\ref{fig:A_Mux}(a) 
and~\ref{fig:A_Mux}(b) we present curves of $\beta$ vs.~$x$ and $\ln[\beta/(1-\beta)]$ 
vs.~$\ln(x)$, respectively. Here, due to the absence of self-loops we observe important differences
with respect to the previous cases: In particular, the curves $\beta$ vs.~$x$ present minima at 
given small values of $x$. This feature can be seen clearer in Fig.~\ref{fig:A_Mux}(b) since it is
magnified there. Also, from Fig.~\ref{fig:A_Mux}(b) we can notice that the range where
$\ln[\beta/(1-\beta)]$ is a linear function of $\ln(x)$ has been shifted upwards for all the values of 
$\alpha$ considered. Therefore, we perform fittings to the curves $\ln[\beta/(1-\beta)]$ vs.~$\ln(x)$
with Eq.~(\ref{betascaling2}) in the interval $\ln[\beta/(1-\beta)]=[0,3.75]$; the bounds of this
interval are marked as dot-dashed lines in Fig.~\ref{fig:A_Mux}(a). The corresponding
values of $\delta$ are reported in the upper inset of Fig.~\ref{fig:A_Mux}(b).
 
Now, under the above conditions, we validate our scaling hypothesis by plotting  $\beta$ vs.~$x^*$ 
and $\ln[\beta/(1-\beta)]$ vs.~$\ln(x^*)$, see Figs.~\ref{fig:A_Mux_collapsed}(a) and 
\ref{fig:A_Mux_collapsed}(b), respectively. 
Remarkably, we observe a clear scaling behaviour also in the unweighted multiplex case (despite
the minima in the curves $\beta$ vs.~$x^*$ for small $x^*$).

\bibliographystyle{plainnat}

\end{document}